\documentstyle[12pt]{article}
\font\bb=msym10
\newcommand{\cn}{\mbox{\bb C}}
\newcommand{\rn}{\mbox{\bb R}}

\newcommand{\p}{\mbox{\bb P}}

\newtheorem{thm}{THEOREM}[section]
\newtheorem{defi}[thm]{DEFINITION}
\newtheorem{lemma}[thm]{LEMMA}
\newtheorem{ex}[thm]{EXAMPLE}
\newtheorem{prop}[thm]{PROPOSITION}
\newtheorem{cor}[thm]{COROLLARY}
\begin{document}

\vspace{35 mm}
{\huge  Projective Quantum Spaces}\\

U. MEYER

{\footnotesize\it D.A.M.T.P.,
University of Cambridge}\\
{\footnotesize\it um102@amtp.cam.ac.uk}

{ \footnotesize September 1994}

\vspace{-5.5cm}
\hfill DAMTP/94-81
\vspace{5.5cm}

{\footnotesize\bf Abstract.}
{\footnotesize Associated to the
standard $SU_{q}(n)$ R-matrices,
we introduce   quantum spheres
$S_{q}^{2n-1}$,
projective quantum spaces
$\cn\p^{n-1}_{q}$, and quantum
Grassmann manifolds
$G_{k}(\cn^{n}_{q})$.
These algebras are shown to be
homogeneous spaces
of standard quantum groups
and are also
quantum principle bundles in the sense of
T. Brzezi\'{n}ski and S. Majid
[1]. }

\section{Introduction}

This paper gives a non-commutative
generalisation of
projective and Grassmann manifolds
in the framework
of quantum groups.
Quantum groups are Hopf algebras
with and additional structure
(quasitriangularity or its
dual version) and can be employed
to construct a
very powerful type of
non-commutative geometry.
For an introduction to quantum
groups and their numerous
other applications
see the review article [4] or the
textbooks [7,2].
We use the  notation and
standard constructions from [4].

The standard  $SU_{q}(n)$ R-matrices
were given in [9]
as
\begin{equation}\label{stan}
R_{n}=q^{-1/n}
\left(q\sum_{i} E_{ii}\otimes E_{ii}
+  \sum_{i\neq j} E_{ii}\otimes E_{jj}
+(q-q^{-1})\sum_{j>i}
 E_{ij}\otimes E_{ji}\right), q\in\rn.
\end{equation}
Here    $E_{ij}$ denote
the elementary  matrices
with components
$(E_{ij})^{a}_{\;b}=
\delta^{a}_{\;i}\delta^{j}_{\;b}.$
These R-matrices are
invertible solutions of
the $n$-dimensional
quantum Yang-Baxter equation and are
of {\it real type}, i.e. obey $
\overline{R}^{\;\;ab}_{n\;\;cd}=
R^{\;\;dc}_{n\;\;ba}$.
Furthermore, let
$\varepsilon_{a_{1}\ldots a_{n}}=
\varepsilon^{a_{1}\ldots a_{n}}
=(-q)^{l(\sigma )}$,
where $l(\sigma )$ is the length
of the permutation
$\sigma (1,\ldots,n)=
(a_{1},\ldots, a_{n})$,
and define
{\it quantum determinants} as
$$det_{q,n}=\nu_{n}^{-1}
\varepsilon_{a_{1}\ldots a_{n}}
 t^{a_{1}}_{\;b_{1}}
\ldots t^{a_{n}}_{\;b_{n}}
\varepsilon^{b_{n}\ldots b_{1}}.$$
The normalisation
factor is   $\nu_{n} =
\varepsilon_{a_{1}\ldots a_{n}}
\varepsilon^{a_{n}\ldots a_{1}}$.

In terms of these data, the
quantum groups $SU_{q}(n)$
were defined in
[9]  as the quotients
$A(R_{n})/_{det_{q,n}=1}$.
This algebra has  well-known antipode $S$
[9, Theorem 4]
and  $\ast$-structure
$t^{a\ast}_{\;\;b}=St^{b}_{\;a}$
such that $tgt^{\dagger }=g$
where $t^{\dagger}=St$ and
$g^{a}_{\;b}=q^{2(a-n)-1}\delta^{a}_{\;b}$.
We also
define quantum groups $U_{q}(n)$
as the quotients of
the algebras $A(R_{n})$
by the relations $det_{q,n}
 det_{q,n}^{\ast} =1$. For $n=1$, one
recovers the undeformed algebra of
complex polynomial functions
on $U(1)$, i.e. $U_{q}(1)$ is the
associative $\cn$-algebra generated
by elements $z$ and $z^{\ast}$ such that
$z^{\ast}=z^{-1}$.

These quantum groups act covariantly on the
algebra of {\it  quantum vectors}
$V(R_{n})$ with
generators $v^{i}, i\in\{1,\ldots,n\}$
and relations $v^{i}v^{j}=
\lambda_{n}^{-1}
R^{\;\;ij}_{n\;\;kl}v^{l}v^{k}$.
Here the normalisation
factors are $\lambda_{n}=q^{1-1/n}$.
The coaction is given by
$v^{i}\mapsto v^{k}\otimes
 t^{\dagger i}_{\;\;\;k}$.
Similarly, the algebra of
{\it quantum covectors}
 $V^{\ast}(R_{n})$ has
generators $x_{i}$, relations
$x_{i}x_{j}=x_{l}x_{k}
\lambda_{n}^{-1}R^{\;\;kl}_{n\;\;ij}
$, and coaction
$x_{i}\mapsto x_{k}\otimes t^{k}_{\;i}$.
In terms of these algebras,
we define
$ \cn_{q}^{n}= V(R_{n})
\underline{\otimes} V^{\ast} (R_{n}) $
as their {\it braided tensor product}
(See [5] for
an introduction to braided geometry).
For the braiding between $ V(R_{n})$ and
$V^{\ast} ( R_{n})$ we
do  not  take the braiding
induced by the coaction
by $SU_{q}(n)$ as given in [6], but
$\Psi(x_{i}\otimes v^{j})
=v^{k}\otimes x_{l}\lambda_{n}
\widetilde{R}^{\;\;lj}_{n\;\;ik}$,
which differs from the
$SU_{q}(n)$-braiding by
a factor
$\lambda_{n}$, i.e.
is the braiding induced by a suitable
dilatonic extension of $SU_{q}(n)$.
We equip this algebra with the
$\ast$-structure $(1\otimes x_{a})^{\ast}=
v^{a}\otimes 1$ and
define the {\it quantum norm}
$N_{q,n}=v^{a}\otimes g_{\;a}^{b}x_{b}$ where
$ g_{\;a}^{b} = \lambda_{n}
\widetilde{R}^{\;\;bc}_{n\;\;ca}$.
The matrix $\widetilde{R}$ is defined as
$((R^{t_{2}})^{-1})^{t_{2}}$, where
$t_{2}$ denotes transposition in the second
tensor component.
Since $R_{n}$ is of real type, $g$ is
symmetric and the
quantum norm is real.
It is known from
[6, Theorem 3.6]  that
for any invertible solution $R$ of the  quantum
Yang-Baxter equation the quantum norm is also
central. For the R-matrices
(\ref{stan}), this general construction
 reproduces the q-metric
introduced above: one finds
$$\widetilde{R}_{n}=
q^{1/n}\left( q^{-1}\sum_{i}
E_{ii}\otimes E_{ii}
+ \sum_{i\neq j} E_{ii}\otimes E_{jj}
-(q-q^{-1})\sum_{j>i} q^{-2(j-i)}E_{ij}
\otimes E_{ji}\right),
$$
and hence $g^{a}_{\;b}= q^{2(a-n)-1}
\delta_{\;a}^{b}.$

We will use the algebras $\cn_{q}^{n}$
to define quantum
spheres and complex projective quantum
spaces. These algebras
are then recovered as base spaces of
{\it quantum principle bundles}
with $U_{q}(n)$ and $SU_{q}(n)$ as
structure groups. The appropriate
definition of a quantum principle
bundle was given by T.
 Brzezi\'{n}ski and S. Majid in
 [1, Definition 4.1]:
an algebra $P$ is a quantum principle
bundle  with structure
quantum group $A$ and base $B$ if (i)
$A$ is a Hopf algebra, (ii)
$P$ is a left $A$-comodule algebra with
coaction $\beta: P\rightarrow
A\otimes P$, and (iii) $B=P^{A} =
\{u\in P: \beta (u) = 1\otimes u\}$
is the subalgebra of $A$-invariant
elements of $P$.
Additionally, there is a freeness and
an exactness condition.
However, if $B$ is a
{\it quantum homogeneous space}, i.e.
if there
is a Hopf algebra
surjection $\pi: P\rightarrow A$
between Hopf algebras $A$
and $P$ and if additionally
\begin{equation}\label{exact}
\ker \pi \subset \cdot
(\ker {\pi}|_{B}\otimes P)
\end{equation}
then the freeness and exactness
condition are automatically satisfied
 [1, Lemma 5.2]
and the quantum homogeneous space
$B=P^{A}$ is a quantum
principle bundle.
In this case the map $\pi$ is
called {\it exact}.

\section{Quantum spheres $S^{2n-1}_{q}$}

Quantum spheres are  quotients
of $\cn_{q}^{n}$ and feature as
base spaces of various
quantum principle bundles.

\begin{defi}
Quantum  spheres
$S^{2n-1}_{q}$ are defined
as the quotients
$\cn_{q}^{n}/_{N_{q,n}=1}.$
\end{defi}

Due to the relation $tgt^{\dagger }=g$, the
quantum spheres
$S^{2n-1}_{q}$ are covariantly coacted
upon by both $U_{q}(n)$ and $SU_{q}(n)$.
For the special cases where $n$ equals
1 or 2, one finds:

\begin{ex}\label{omm}
$ \;S^{1}_{q}\cong U_{q}(1)$ and
$S^{3}_{q}\cong SU_{q}(2) $ as
$\ast$-algebras.
\end{ex}

{\it Proof.}
The first isomorphism is trivial
and the second one is
easily established by
rescaling of the generators of
 $S_{q}^{3}$. This isomorphism
only holds
if one introduces the
normalisation factor
$\lambda_{2}$ in the definition of the
braiding in $\cn_{q}^{2}$.
\rule{2mm}{2mm}

\begin{prop}\label{Kugel}
There are $n$  $\ast$-algebra injections
$\imath_{m}:S^{2n-1}_{q}
\hookrightarrow U_{q}(n)$.
\end{prop}

{\it Proof.}
In this proof,  there is  no
summation over the variable   $m$.
Define the maps $\imath_{m}$ by
$x_{a}\mapsto t^{m}_{\;a}$.
By inspection of (\ref{stan}) one finds
$R^{\;\;mm}_{n\;\;cd}=q^{1-1/n}
\delta^{m}_{\;c}\delta^{m}_{\;d}$ and
hence the relations in $A(R_{n})$ imply
$q^{1-1/n} t^{m}_{\;a}t^{m}_{\;b}=
 R^{\;\;mm}_{n\;\;cd}t^{c}_{\;a}t^{d}_{\;b}=
t^{m}_{\;d}t^{m}_{\;c}R^{\;\;cd}_{n\;\;ab},$
i.e.  the generators
$t^{m}_{\;a}, a\in\{1,\ldots, n\}$
obey the relations of
$V^{\ast}(R_{n})$. Since $R_{n}$
is of real type,
the generators
$t^{\dagger a}_{\;\;\;m},
a\in\{1,\ldots, n\}$ then
obey  the
 $V( R_{n})$-relations.
Next note that the  relations in $A(R_{n})$
can be written as
$t^{\dagger a}_{\;\;\; b}
 R^{\;\;cb}_{n\;\;de}t^{d}_{\;g}
=t^{c}_{\;d} R^{\;\;da}_{n\;\;gb}
t^{\dagger b}_{\; \;\;e}$
and hence with $R^{\;\;am}_{n\;\;mg}= q^{1-1/n}
 \delta^{a}_{\;m}\delta^{m}_{\;g}$
one finds
$t^{m}_{\;i}t^{\dagger j}_{\;\;\;m}=
\widetilde{R}^{\;\;dj}_{n\;\;if}
t^{\dagger f}_{\;b}
R^{\;\;mb}_{n\;\;cm}t^{c}_{\;d}=
q^{1-1/n}
\widetilde{R}^{\;\;dj}_{n\;\;if}
t^{\dagger f}_{\;m}t^{m}_{\;d}$.
Thus the  relations between $t^{m}_{\;i}$ and
$t^{\dagger j}_{\;\;\;m}$ reproduce
the braiding between $V( R_{n})$ and
$V^{\ast}( R_{n})$ given above.
This is the place where one
 needs the normalisation factor
$\lambda_{n}$ in the definition
of the braiding.
Finally note,  that
$t^{\dagger a}_{\;\;\;m}g^{b}_{\;a}t^{m}_{\;b}
=
\lambda_{n} t^{\dagger a}_{\;\;\;m}
\widetilde{R}^{\;\;bi}_{n\;\;ia}
t^{m}_{\;b}
=
t^{m}_{\;i}t^{\dagger i}_{\;\;\;m}
=1$. Hence
the maps $\imath_{m}$ are
$\ast$-algebra injections.
\rule{2mm}{2mm}

For the construction
of quantum principle bundles over
these quantum spheres we need the
following lemma:

\begin{lemma}\label{rtrt}
There are surjective $\ast$-Hopf
algebra morphisms
$$
\begin{array}{rccl}
\alpha_{n}:& U_{q}(n)& \rightarrow
& U_{q}(n-1)\\
\gamma_{n}:&  SU_{q}(n)  &
\rightarrow & SU_{q}(n-1).
\end{array}
 $$
\end{lemma}

{\it Proof.}
By inspection of (\ref{stan})
 one finds that
for $i,j >1 $, $R^{\;\;ij}_{n\;\;\;kl}=
R^{\;\;\;(i-1)(j-1)}_{n-1\;\;(k-1)(l-1)}$
and also that
$R^{\;\;11}_{n\;\;ij}=
q^{1-1/n}\delta^{1}_{\;i}\delta^{1}_{\;j}$.
Thus  the elements
$t^{i}_{\;j}; i,j\in\{2,\ldots, n \}$ of
$A(R_{n})$ obey the
relations of $A(R_{n-1})$ and the
quotient maps by the ideal
generated by $t^{1}_{\;1}=1$
and
$t^{1}_{\;i}=t^{i}_{\;1}=0,
i\in\{2,\ldots, n \}$
 are surjections
$ A(R_{n})\rightarrow A(R_{n-1})$.
These quotient maps also have the
property
$det_{q,n}\mapsto  det_{q,n-1}$
and thus descend
to surjective $\ast$-Hopf algebra maps
$  U_{q}(n)\rightarrow U_{q}(n-1)$
and $SU_{q}(n)\rightarrow SU_{q}(n-1)$,
 which we denote by
$\alpha_{n}$ and $\gamma_{n}$,
 respectively.
\rule{2mm}{2mm}

\begin{thm}\label{lala}
The Hopf algebra $U_{q}(n)$ is a quantum
principle bundle with
base  $S^{2n-1}_{q}$ and structure
quantum group $U_{q}(n-1)$:
\begin{equation}\label{hom}
S^{2n-1}_{q} \cong U_{q}(n)^{U_{q}(n-1)}
\end{equation}
 The left covariant coaction of
$U_{q}(n-1)$
on $U_{q}(n)$
is given by $(\alpha_{n}\otimes id)
\circ\Delta$
in terms of the morphism
$\alpha_{n}$ from lemma \ref{rtrt}.
\end{thm}

{\it Proof.}
The subalgebra of $U_{q}(n-1)$-invariant
elements of $U_{q}(n)$
is generated  by  $t^{1}_{\;i},
i\in\{1,\ldots, n\}$
and their conjugates.
Proposition \ref{Kugel} then implies
(\ref{hom}).
Since $\ker \alpha_{n} = \ker
\alpha_{n}|_{S^{2n-1}_{q}}$
the  maps $\alpha_{n}$ are
exact in the sense of (\ref{exact})
and the quantum homogeneous
space is a quantum principle bundle.
\rule{2mm}{2mm}

One can repeat this
construction with the
maps $\gamma_{n}: SU_{q}(n)
\rightarrow SU_{q}(n)$
from lemma \ref{rtrt} to
establish that $SU_{q}(n)$
is also a  quantum principle bundle
over a quantum sphere:

\begin{thm}
The Hopf algebra  $SU_{q}(n)$ is a quantum
 principle bundle
with base
$S_{q}^{2n-1}$ and structure
quantum group $SU_{q}(n-1)$:
$$S_{q}^{2n-1}\cong SU_{q}(n)^{SU_{q}(n-1)}$$
The left covariant coaction of
$SU_{q}(n-1)$ on
$SU_{q}(n)$ is given by
$( \gamma_{n}\otimes id)\circ
\Delta$.
\end{thm}

\section{Complex projective quantum spaces}

We now turn our attention to
complex projective spaces
which are defined as
$U_{q}(1)$-covariant subalgebras
of  quantum spheres
 $S^{2n-1}_{q}$. For this purpose
note that
by virtue of lemma \ref{rtrt},
the right covariant
$\ast$-coaction
$\beta: \cn^{n}_{q}\rightarrow
\cn^{n}_{q}\otimes U_{q}(n)$
descends to a covariant
$U_{q}(1)$-coaction defined as
$(id\otimes (\alpha_{2}\circ
\alpha_{3}\circ\ldots
\circ \alpha_{n}))
\circ\beta$. This coaction is
given explicitly by
$x^{i}\mapsto x^{i}\otimes z$ and
also descends
to a covariant $\ast$-coaction on
 $S^{2n-1}_{q}$.

\begin{defi}\label{prodet}
Complex projective quantum space
 $\cn\p_{q}^{n}$ is defined as
$$ \cn\p_{q}^{n-1} =
(S^{2n-1}_{q})^{U_{q}(1)}, $$
the subalgebra of $U_{q}(1)$-invariant
elements of the
quantum sphere $S^{2n-1}_{q}$.
\end{defi}

Since quantum projective spaces are merely
subalgebras of
quantum spheres $S^{2n-1}_{q}$, they are
also covariantly coacted
upon by $SU_{q}(n)$ and $U_{q}(n)$. Note
also that their relations
are independent of the
normalisation factor $\lambda_{n}$ which we
introduced in the
definition of $\cn_{q}^{n}$, i.e. at this
stage one could work
with the standard $SU_{q}(n)$-braiding.

For the construction  of
quantum principle bundles on these
complex projective
quantum spaces
 we need the following lemma:

\begin{lemma}
There are a surjective $\ast$-Hopf
algebra morphisms
$$\delta_{n}: SU_{q}(n)
\rightarrow U_{q}(n-1).$$
\end{lemma}

{\it Proof.}
The proof is  similar to  the proof
of lemma \ref{rtrt}.
The maps $\delta_{n}$ are defined like
$\beta_{n}$ with the only
difference that $\delta_{n}(t^{1}_{\;1})
= (det_{q,n-1})^{\ast}$.
\rule{2mm}{2mm}

\begin{thm}\label{piet}
The Hopf algebra $SU_{q}(n)$ is a
quantum
 principle bundle
with base
$\cn\p^{n-1}_{q}$ and structure
quantum group $U_{q}(n-1)$:
$$\cn\p^{n-1}_{q}\cong
SU_{q}(n)^{U_{q}(n-1)}.$$
The coaction of $U_{q}(n-1)$ on
$SU_{q}(n)$ is given by
$(\delta_{n}\otimes id)\circ\Delta$.
\end{thm}

{\it Proof.}
In this case,  invariant
generators are products
$t^{\dagger i}_{\;\;\;1}t^{1}_{\;j},
i,j\in\{1, \ldots, n\}$
which generate the subalgebra
of $U_{q}(1)$-invariant elements of
$S^{2n-1}_{q}$,
i.e. $\cn\p^{n-1}_{q}$.
The kernel of   $\delta_{n}$
is generated by $t^{i}_{\;1}$ and
$t^{1}_{\;i}$, where
$i\in\{2,\ldots n\}$ and  the kernel
 of the restricted map
$\ker\delta_{n}|_{\cn\p^{n-1}_{q}}$ has
generators $t^{1}_{\;i}
t^{\dagger j}_{\;\;\;1}$ where
either $i$ or
$j$ is larger that 1.
The exactness of $\delta_{n}$
is then established
by the trivial observation
$t^{\dagger i}_{\;\;\;1}=\cdot
(t^{\dagger i}_{\;\;\;1}
t^{1}_{\;j}\otimes
t^{\dagger j}_{\;\;\;1})$
and the  relation
$t^{1}_{\;i} =
t^{1}_{\;i}t^{\dagger j}_{\;\;\;1}
g^{k}_{\;j}t^{1}_{\;k}
=\lambda_{n}
\widetilde{R}^{\;\;dj}_{n\;\;if}
t^{\dagger f}_{\;\;\;1}t^{1}_{\;d}
g^{k}_{\;j}t^{1}_{\;k}
=\cdot (t^{\dagger f}_{\;\;\;1}t^{1}_{\;d}
\otimes \lambda_{n}
\widetilde{R}^{\;\;dj}_{n\;\;if}
g^{k}_{\;j}t^{1}_{\;k})$.
Here we used
the formulae from the  proof of
proposition
\ref{Kugel} and the fact
that for $i\neq 1$,
$\widetilde{R}^{\;\;dj}_{n\;\;if}$ is
non-zero only if $d\neq 1$.
\rule{2mm}{2mm}

As a corollary of theorem \ref{lala}
one finds with definition
\ref{prodet} that
complex projective quantum
spaces $\cn\p_{q}^{n}$
can also be regarded as
homogeneous
quantum spaces of $U_{q}(n)$:

\begin{thm}\label{specc}
The Hopf algebra $U_{q}(n)$ is a
quantum principle bundle
with base $\cn\p^{n}_{q}$
and structure quantum group $U_{q}(1)
\otimes U_{q}(n-1)$:
$$\cn\p^{n}_{q} \cong U_{q}(n)^{U_{q}(1)
 \otimes U_{q}(n-1)}$$
\end{thm}

The covariant coaction by $U_{q}(1)
 \otimes U_{q}(n-1)$ is
given explicitly in the
next section, where we
generalise this construction to
quantum Grassmann manifolds.

We now restrict our attention
to the case
of $n=2$. In this case, the
algebra $SU_{q}(2)^{U_{q}(1)}$
is a deformation of the two-sphere
and was already discussed by
P. Podle\`{s} in [8],
where it was denoted by $S_{q}^{2}$.
As a corollary of theorem
\ref{piet} we hence find:

\begin{cor}
Projective quantum space
$\cn\p^{1}_{q}$ is isomorphic
to Podle\`{s}'s quantum sphere
$S^{2}_{q}$.
\end{cor}

The $n=2$  case
is also in the undeformed case quite
 remarkable since   all
spaces involved
are spheres:
$SU(2)\cong S^{3}$, $U(1)\cong S^{1}$
and $\cn\p^{1}\cong S^{2}$.
The corresponding principle bundle
is the celebrated Hopf
fibration of the three sphere.
In the quantum case one finds with
example \ref{omm} and theorem
 \ref{piet}:

\begin{cor} {\bf ( Hopf fibration of
the quantum 3-sphere)}
There is a quantum principle bundle
\begin{equation}\label{hopf}
S^{2}_{q}\cong (S_{q}^{3})^{S^{1}_{q}}
\end{equation}
\end{cor}

The quantum homogeneous space (\ref{hopf})
was already
discussed in [11], although not as quantum
principle bundles, an appropriate definition
of which was
only given a few years later.

\section{Quantum Grassmann manifolds}

The construction encountered in
theorem \ref{specc}
can   be generalised to give a
definition of quantum Grassmann
manifolds as
quantum principle bundles. For
this purpose,
we prove the following lemma:

\begin{lemma}\label{lulu}
There are surjective  $\ast$-Hopf algebra
morphisms
$$\alpha_{k,n}: U_{q}(n)\rightarrow
U_{q}(k)\otimes U_{q}(n-k).$$
\end{lemma}

{\it Proof.}
Note that for $i,j<n$,
$R^{\;\;ij}_{n\;\;kl}=
R^{\;\;\;ij}_{n-1\;kl}$,
i.e. the elements $t^{i}_{\;j};
 i,j\in\{1,\ldots,n-1\}$
of $A(R_{n})$ obey the
relations of $A(R_{n-1})$. Similar
to lemma \ref{rtrt} one can thus define
surjective $\ast$-Hopf algebra
morphisms $\alpha^{\prime}_{n}:
U_{q}(n)\rightarrow U_{q}(n-1)$
by dividing by
$t^{n}_{\;n}=1$ (no summation!) and
$t^{n}_{\;i}=t^{i}_{\;n}=0,
i\in\{1,\ldots, n-1\}$. The maps
$ (
(\alpha_{k+1}\circ\alpha_{k+2}\circ
\ldots\circ\alpha_{n})\otimes
(\alpha_{n-k+1}^{\prime}\circ\ldots
\circ\alpha^{\prime}_{n}))
\circ\Delta$
are then the required $\ast$-Hopf algebra
maps $\alpha_{k,n}$.
\rule{2mm}{2mm}

The covariant coaction  by $U_{q}(1)
 \otimes U_{q}(n-1)$ on $U_{q}(n)$
from theorem \ref{specc}
is  given by
$(\alpha_{1,n}\otimes  id)\circ\Delta.$
However, the real merit of this
lemma is that it leads to a straightforward
generalisation
from quantum projective spaces
to quantum Grassmann manifolds:

\begin{defi}\label{graa}
Quantum Grassmann manifolds are
defined as the quantum
principle bundles
$$
G_{k}(\cn_{q}^{n}) =
U_{q}(n)^{U_{q}(k)\otimes U_{q}(n-k)}
$$
where the covariant coaction
is given by $(\alpha_{k,n}\otimes id)
\circ\Delta$. This definition includes the
case of projective quantum spaces as
 $G_{1}(\cn_{q}^{n})
 = \cn\p^{n}_{q}$.
\end{defi}

A useful set of generators of
 $G_{k}(\cn^{n}_{q})$ is given by the elements
$$(\varepsilon_{c_{1}\ldots c_{k}}
t^{c_{1}}_{\;a_{1}}\ldots t^{c_{k}}_{\;a_{k}}
 )^{\ast}
 (
\varepsilon_{d_{1}\ldots d_{k}}
t^{d_{1}}_{\;b_{1}}\ldots t^{d_{k}}_{\;b_{k}}
 ),$$
where $c_{i},d_{i} \in \{ 1,\ldots, k\}$ and
$ a_{i},b_{i}\in\{1,\ldots,n\}$.

An earlier paper by E. Taft and
 J. Towber [10] (see also [3])
presented a related approach to
quantum Grassmann manifolds and
 studied the algebra generated by
elements
$\varepsilon_{d_{1}\ldots d_{k}}
t^{d_{1}}_{\;b_{1}}\ldots
t^{d_{k}}_{\;b_{k}}$,
where the $t^{a}_{\; b}$
are the generators of
$A(R_{n})$ and not of the quotient
$U_{q}(n)$. Thus a quotient
of a subalgebra
of the algebra from [10, Proposition 3.4]
is
the quantum Grassmann manifold
 $G_{k}(\cn_{q}^{n})$.

\pagebreak
{\bf \LARGE \bf References}

{\footnotesize

\begin{tabbing}

XX\= \kill \\

1.\>
T.~Brzezinski and S.~Majid,
{\em Comm. Math. Phys.} 157, 591 (1993).\\

2.\>
V.~Chari and A.~Pressley,
{\em A Guide to Quantum Groups},
 C.U.P., 1994.\\

3.\> V. Lakshmibai and N. Reshetikhin,
{\em C. R. Acad. Sci. Paris} I
 313 (3), 121 (1991)\\

4.\>
S.~Majid,
{\em Int. J. Mod. Phys. A} 5, 1  (1990).\\

5.\>
S.~Majid,
 in: M.-L. Ge and H.~J. {de Vega},
{\em Proc. 5th Nankai Workshop},
World Scientific,  1992.\\

6.\>
S.~Majid, {\em J. Math. Phys.}
 34, 1176 (1993).\\

7. \>
S.~Majid, {\em Foundations of Quantum
Group Theory}, C.U.P. (to be published)\\

8.\>
P.~Podles, {\em Lett. Math. Phys.}
14, 193 (1987).\\

9.\>
N. Reshetikhin, L. Takhtadzhyan,
and L.  Faddeev,
{\em Leningrad Math. J.} 1 (1), 193 (1990).\\

10.\>
E.~Taft and J.~Towber, {\em J. Algebra}
142, 1 (1991).\\

11.\>
S.~L. Woronowicz, {\em Publ. RIMS }
23, 117 (1987).

\end{tabbing}}

\end{document}